\documentclass[preprint2]{aastex63}

\usepackage[flushleft]{threeparttable}
\usepackage{CJK}

\shorttitle{Binary in C1-Sa}

\begin{document}
\begin{CJK*}{UTF8}{gbsn}

\title{Binary Formation in a 100 $\mu$m-dark Massive Core}

\author[0000-0002-8469-2029]{Shuo Kong (孔朔)}
\affil{Steward Observatory, University of Arizona, Tucson, AZ 85719, USA}

\author[0000-0001-5653-7817]{H\'ector G. Arce}
\affil{Department of Astronomy, Yale University, New Haven, CT 06511, USA}

\author[0000-0002-6195-0152]{John J. Tobin}
\affil{National Radio Astronomy Observatory, 520 Edgemont Rd., Charlottesville, VA 22903, USA}

\author[0000-0001-7511-0034]{Yichen Zhang}
\affiliation{Department of Astronomy, University of Virginia, Charlottesville, VA 22904-4325, USA}
\affiliation{RIKEN Cluster for Pioneering Research, Wako, Saitama 351-0198, Japan}

\author[0000-0002-7026-8163]{Mar\'ia Jos\'e Maureira}
\affil{Max-Planck-Institute for Extraterrestrial Physics (MPE), Giessenbachstr. 1, D-85748 Garching, Germany}

\author[0000-0001-5253-1338]{Kaitlin M. Kratter}
\affil{Steward Observatory, University of Arizona, Tucson, AZ 85719, USA}

\author[0000-0003-2133-4862]{Thushara G.S. Pillai}
\affil{99 Millstone Road, MIT Haystack Observatory, Westford, MA, 01186, USA}

\begin{abstract}
We report high-resolution ALMA observations toward a massive protostellar core C1-Sa ($\sim$30 M$_\odot$) in the Dragon Infrared Dark Cloud. At the resolution of 140 AU, the core fragments into two kernels (C1-Sa1 and C1-Sa2) with a projected separation of $\sim$1400 AU along the elongation of C1-Sa, consistent with a Jeans length scale of $\sim$1100 AU.  Radiative transfer modeling using RADEX indicates that the protostellar kernel C1-Sa1 has a temperature of $\sim$75 K and a mass of 0.55 M$_\odot$. C1-Sa1 also likely drives two bipolar outflows, one being parallel to the plane-of-the-sky. C1-Sa2 is not detected in line emission and does not show any  outflow activity but exhibits ortho-H$_2$D$^+$ and N$_2$D$^+$ emission in its vicinity, thus it is likely still starless. Assuming a 20 K temperature, C1-Sa2 has a mass of 1.6 M$_\odot$. At a higher resolution of 96 AU, C1-Sa1 begins to show an irregular shape at the periphery, but no clear sign of multiple objects or disks. We suspect that C1-Sa1 hosts a tight binary with inclined disks and outflows. Currently, one member of the binary is actively accreting while the accretion in the other is significantly reduced. C1-Sa2 shows hints of fragmentation into two sub-kernels with similar masses, which requires further confirmation with higher sensitivity. 
\end{abstract}


\section{Introduction}
\end{CJK*}

Binary formation is an essential part of star formation,
as young stellar objects usually come in pairs 
\citep[or multiple systems, see][]{2014prpl.conf..267R,2022arXiv220310066O}.
In particular, massive stars (M $\ga$ 8 M$_\odot$)
have a remarkably high multiplicity frequency 
\citep[approaching 100\%, see][]{2012MNRAS.424.1925C,2017ApJS..230...15M}, 
but their formation is still not well understood \citep{2018ARA&A..56...41M}.
Investigating the formation of massive binaries is therefore
pivotal for a better understanding of massive star formation.
However, statistical studies of high-mass protostars
are quite challenging because the sample
size is typically much smaller than that of low-mass protostars.
In addition, massive protostars are typically in crowded
clusters deeply embedded in molecular clouds 
\citep{2007ARA&A..45..481Z}. Observations in such regions
are challenging for optical and infrared wavelengths. Statistically
meaningful surveys have only started to come out within the last decade
\citep[e.g.,][]{2013A&A...550A.107S}. 

Thus far, studies of massive binary stars have not
put much emphasis on the early phase \citep{2013ARA&A..51..269D}.
However, the early phase is extremely important 
because the binary properties we see in later stages may be very
different from that during the formation process. 
Therefore, to fully understand the ``true'' origin of massive binaries,
it is crucial to focus on the very early stages
of the binary formation. In particular, information about 
the initial separation between the binary,
the mass ratio of the binary, and potentially the relative
orientation of the accretion disks will be helpful for testing
massive binary formation theories.

Several observations have revealed massive binary formation
at the protostellar phase \citep[e.g.,][]{2016A&A...593A..49B,2020ApJ...900L...2T}. Most recently,
\citet{2022ApJ...931L..31C} observed a (proto)binary in 
a massive core that was previously thought to be prestellar,
making it one of the earliest-stage forming binaries that has
the potential to become massive. Such observations put stringent
constraints on massive binary formation theories 
\citep[e.g.,][]{2009Sci...323..754K,2020A&A...644A..41O,2021A&A...652A..69M}

Star formation happens in dense molecular cores
\citep{2007ARA&A..45..339B}. These cores provide laboratories
for observing the earliest phase of binary formation.
Due to the high column density toward star-forming regions
and the relatively low temperature in molecular clouds,
protostellar cores are best detectable at mm wavelengths
(e.g.,\citealt{2019ApJ...873...31K};
\citealt{2019ApJ...886..102S}; \citealt{2022A&A...662A...8M}).
For distant high-mass star-forming regions, ALMA is 
the best choice with its superb resolution and sensitivity.
For instance, \citet{2019NatAs...3..517Z} used 
ALMA to probe a forming binary in IRAS 07299-1651 which is
a massive star-forming region that is 1.7 kpc away.
With a resolution of $\sim$0.03\arcsec\ or $\sim$50 AU 
resolution, they found that the IRAS 07299-1651
binary protostar appears to be separated by $\sim$200 AU, which was
interpreted as due to disk fragmentation. Additional ALMA
observations that zoom into the heart of massive protostellar 
cores will allow future statistical studies of massive binary 
formation, augmenting surveys at shorter wavelengths
\citep[see, e.g., ][]{2018A&A...620A.116G} and 
revolutionizing our understandings of massive star formation.

\citet[][hereafter K18]{2018ApJ...867...94K} observed a 
massive protostellar core C1-Sa ($\sim$30 M$_\odot$) and an
accompanying low-mass protostellar core C1-Sb ($\sim$2 M$_\odot$)
in the Dragon infrared dark cloud 
\citep[Dragon IRDC,][]{2018RNAAS...2...52W,2021ApJ...912..156K} 
with ALMA \citep[also see][hereafter T16]{2016ApJ...821L...3T}.
The Dragon IRDC (aka G28.34+0.06 or G28.37+0.07)
is at a distance of 4.8 kpc\footnote{We adopt the distance for consistency and comparison with previous literature, see \S\ref{sec:dist} for details.},
and the cores are embedded in a
region that is dark at up to 100 $\mu$m \citep{2012A&A...547A..49R},
indicating that it is at the earliest phase of star formation.
With a 0.2\arcsec\ spatial resolution ($\sim$960 AU at a distance of
4.8 kpc), the K18 1.3 mm continuum image showed an elongated structure
in the center of C1-Sa, which hinted at the existence of two fragments
(see their Figure 1) whose separation is $\sim$0.2\arcsec,
i.e., the resolution limit of their ALMA data. 
In addition, C1-Sa likely drives two bipolar outflows.
The first (hereafter C1-Sa outflow1) was identified in T16.
The second (hereafter C1-Sa outflow2) 
was identified as a filament in T16
because the elongated structure showed up in CO channel maps
contemporaneously between 74 and 80 km s$^{-1}$, i.e., it showed no
blue- or red-shifted lobes. However, based on the data
from \citet{2019ApJ...874..104K}, the elongated CO structure 
showed well-matching SiO counterparts, 
indicative of a bipolar outflow, 
which motivated this study.
The suspected binary fragmentation, if confirmed, would provide
an excellent example of massive binary formation at the earliest stage.

In this paper, we present ALMA Cycle 7 high-resolution 
observations toward C1-Sa. We show that the core is indeed
a binary system, and analyze the ALMA data to characterize its
physical properties. To our knowledge, the C1-Sa binary system
is the most distant proto-binary at the earliest-stage
which also has the potential to become massive.
It shows the very beginning of massive binary formation in an IRDC.
In the following, \S\ref{sec:obsalma}, we introduce the ALMA observations and 
data reduction. Next, we report our
findings about the binary system in \S\ref{sec:results}. 
We  discuss the implications of our results in \S\ref{sec:disc}.
Finally, \S\ref{sec:conc} summarizes and concludes the paper.

\section{ALMA observations and data reduction}\label{sec:obsalma}

In this paper, we utilize multiple ALMA datasets that cover the cores. 
First, ALMA Cycle 2 project 2013.1.00248.S 
(hereafter {\tt C2}) observed the area with a 
$\sim$0.2\arcsec\ resolution. 
It includes the 1.3 mm continuum
and  CO(2-1) line, an outflow tracer. 
Details of the {\tt C2} data are present in T16 and K18.
Second, ALMA Cycle 3 project 2015.1.00183.S 
(hereafter {\tt C3}) mosaicked the main body of the Dragon
\citep[see Figure 1 in][]{2019ApJ...874..104K} with the 1.3 mm
continuum and the outflow tracers CO(2-1) and SiO(5-4).
Details of the {\tt C3} data are included in 
\citet{2019ApJ...873...31K}
and \citet{2019ApJ...874..104K}.
Third, ALMA Cycle 4 project 2016.1.00988.S 
(hereafter {\tt C4}) observed the region
in ortho-H$_2$D$^+$ (J$_{\rm Ka, Kc}$ = 1$_{1, 0}$-1$_{1, 1}$).
The {\tt C4} data will be
presented in detail in a future paper. The ortho-H$_2$D$^+$
line data shown in this paper has a synthesized beam of
$\sim1\arcsec$ and a sensitivity of 0.015 Jy beam$^{-1}$
per 0.2 km s$^{-1}$ channel.
Finally, the  ALMA Cycle 7 project 2019.1.00255.S 
(hereafter {\tt C7}) observed
the C1-Sa area with a $\sim$0.03\arcsec\ synthesized beam.

Here, we report the high-resolution {\tt C7} band 6 continuum and molecular line observations. A dedicated 2 GHz baseband was used for the continuum observation. The baseband was centered at 234 GHz (1.3 mm) with a channel width of 1.1 MHz. Line contamination was checked in visibilities and excluded from imaging. The continuum observation served the main goal of detecting the core fragmentation via dust emission. A spectral window for CO(2-1) was included with a 0.18 km s$^{-1}$ channel width. The line was mainly used to detect the CO outflows from the fragments in the massive C1-Sa core. We also included spectral windows for $^{13}$CO(2-1) and C$^{18}$O(2-1) to probe the kinematics in the core. Both lines were set to a 0.19 km s$^{-1}$ spectral resolution. The rest of the spectral windows were set for CH$_3$CN v=0 12(1)-11(1), F=11-10 (channel width 0.18 km s$^{-1}$), H(30)$\alpha$ (channel width 0.18 km s$^{-1}$), SO 3$\Sigma$ v=0 6(5)-5(4) (channel width 0.18 km s$^{-1}$), SiO(5-4) (channel width 0.39 km s$^{-1}$), CH$_3$OH 4(2)-3(1) E1 vt=0 (channel width 0.39 km s$^{-1}$), H$_2$CO 3(0,3)-2(0,2) (channel width 0.39 km s$^{-1}$), H$_2$CO 3(2,1)-2(2,0) (channel width 0.39 km s$^{-1}$), H$_2$CO 3(2,2)-2(2,1) (channel width 0.39 km s$^{-1}$).

\begin{deluxetable*}{lccrccc}
\tablecaption{{\tt C7} Data Imaging Products \label{tab:imaging}}
\tablehead{
\colhead{Product ID} &
\colhead{Weighting} &
\colhead{Beamsize} &
\colhead{PA} &
\colhead{RMS$_{\rm cont}$} &
\colhead{RMS$_{\rm line}$} &
\colhead{Figure}\\
\colhead{} &
\colhead{(robust)} &
\colhead{} &
\colhead{} &
\colhead{$\mu$Jy beam$^{-1}$} &
\colhead{mJy beam$^{-1}$} &
\colhead{}
}
\startdata
{\tt C7Taper} & uv-tapering & $0.14\arcsec\times0.11\arcsec$ & -65\arcdeg & - & 1.8 & \ref{fig:showout} \\
{\tt C7Natural} & Natural & $0.077\arcsec\times0.052\arcsec$ & -61\arcdeg & - & 1.6 & \ref{fig:chanco} \\
{\tt C7Briggs1} & Briggs(1) & $0.064\arcsec\times0.044\arcsec$ & -48\arcdeg & - & 1.7 & \ref{fig:pv} \\
{\tt C7Briggs0.5}\tablenotemark{a} & Briggs(0.5) & $0.038\arcsec\times0.030\arcsec$ & -43\arcdeg & 25 & 1.8 & \ref{fig:showbinary},\ref{fig:showout},\ref{fig:spec} \\
{\tt C7Briggs-0.5} & Briggs(-0.5) & $0.022\arcsec\times0.020\arcsec$ & 82\arcdeg & 50 & - & \ref{fig:mf} \\
{\tt C7Briggs-1} & Briggs(-1) & $0.022\arcsec\times0.018\arcsec$ & 72\arcdeg & 70 & - & \ref{fig:mf}
\enddata
\tablecomments{RMS$_{\rm cont}$ is the continuum RMS noise. A dash sign ``-'' indicates that the data is not used. RMS$_{\rm line}$ is the line cube RMS noise per 0.2 km s$^{-1}$ channel measured in emission-free channels.}
\tablenotetext{a}{The fiducial data, see \S\ref{sec:obsalma}.}
\end{deluxetable*}

The {\tt C7} ALMA observations included two 12m-array configurations. The first was the C43-6 configuration with an angular resolution of $\sim$0.15\arcsec. The second was the C43-9 configuration with an angular resolution of $\sim$0.025\arcsec. The C43-6 scheduling block was executed on 2021-07-11. It used the bandpass calibrator J1924-2914 and the phase calibrator J1851+0035. The total on-source time was 22 min. The C43-9 scheduling block was executed on 2021-09-25 and 2021-09-26. Both executions used the bandpass calibrator J1924-2914 and the phase calibrator J1834+0301. The total on-source time was 99 min. In the following, all {\tt C7} imaging products are made using both the C43-6 and C43-9 data.

For imaging, we adopt a few different parameters that lead to different versions of the final image product. The purpose is to examine the structures at different scales. The fiducial version uses the Briggs weighting scheme with a robust number of 0.5 (hereafter referred to as {\tt C7Briggs0.5}). The continuum root-mean-square noise (RMS$_{\rm cont}$) is 25 $\mu$Jy beam$^{-1}$. Then, to improve the signal-to-noise ratio (SNR), we also use a robust number of 1 (hereafter {\tt C7Briggs1}), a natural weighting scheme (hereafter {\tt C7Natural}) and a uv-tapering scheme (at 1500 k$\lambda$ with $\lambda\sim1.3$ mm, hereafter {\tt C7Taper}). Finally, to have the best spatial resolution, we also use a Briggs weighting with a robust number of -0.5 (hereafter {\tt C7Briggs-0.5}) and -1 (hereafter {\tt C7Briggs-1}). These imaging products are summarized in Table \ref{tab:imaging} along with the synthesized beam and the figures in which they are shown.

\section{Results and Analyses}\label{sec:results}

\subsection{C1-Sa Core Fragmentation}\label{subsec:frag}

\begin{figure*}[htb!]
\centering
\epsscale{1.1}
\plotone{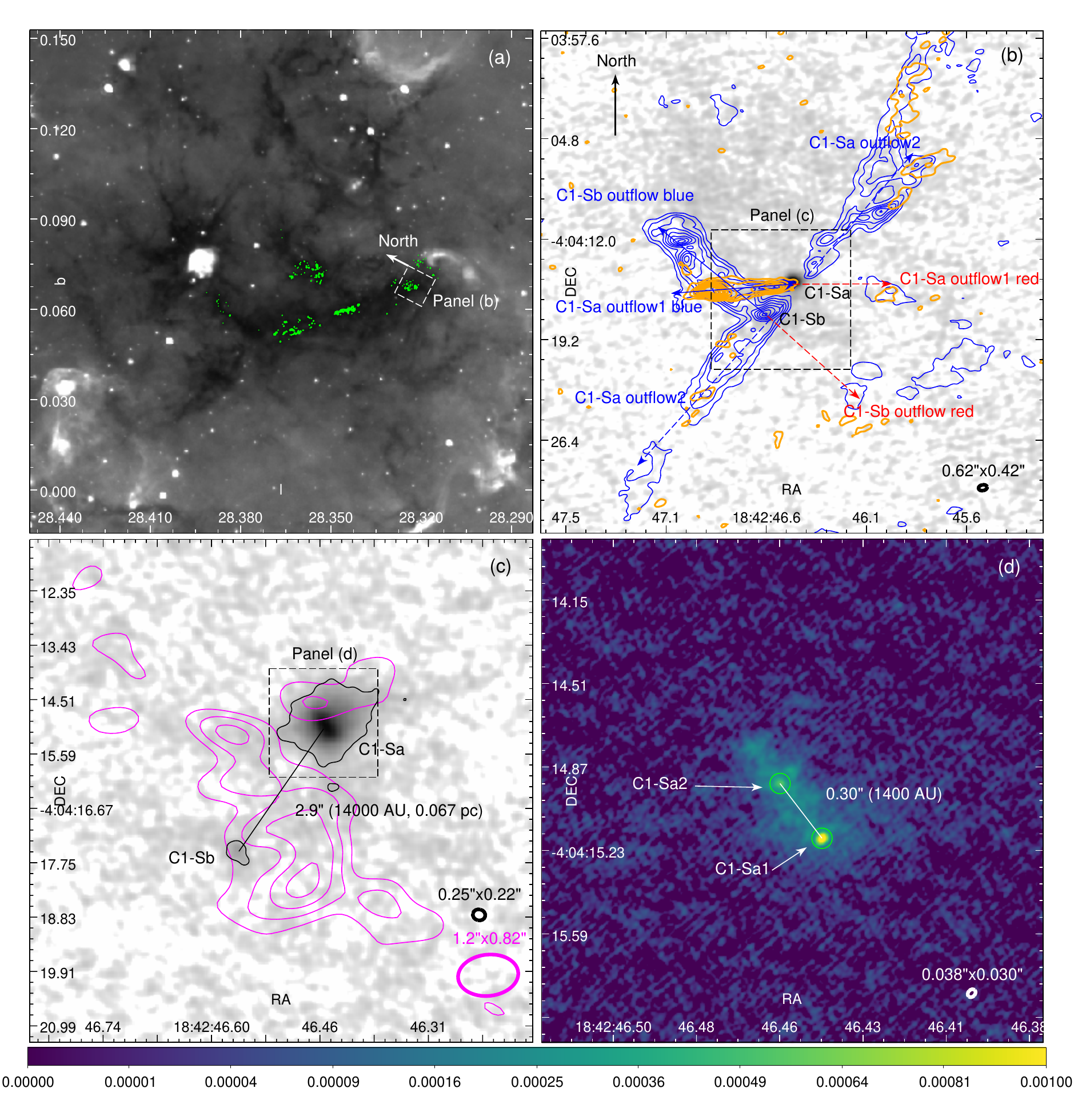}
\vspace{-10pt}
\caption{
{\bf (a):} Spitzer $8\mu$m image of the Dragon IRDC (in galactic coordinates). The green contours mark the 1.3 mm continuum dust emission from {\tt C3} (see \S\ref{sec:obsalma}) at SNR=5, 10, 15 (RMS$_{\rm cont}$=80 $\mu$Jy beam$^{-1}$). 
{\bf (b):} A zoom-in view of the C1-Sa and C1-Sb cores (with the {\tt C3} data, in equatorial coordinates, black ellipse beam). The grey scale map shows the 1.3 mm continuum emission from panel (a). The blue contours show the integrated intensity of $^{12}$CO(2-1) for C1-Sa outflow2 at SNR=9, 18, ..., 72 (RMS$_{\rm mom0}$=10 mJy beam$^{-1}$ km s$^{-1}$). The orange contours show the integrated intensity of SiO(5-4) at SNR=3, 7, ..., 31 (RMS$_{\rm mom0}$=6 mJy beam$^{-1}$ km s$^{-1}$).
{\bf (c):} A zoom-in view of C1-Sa and C1-Sb, showing the {\tt C2} 1.3 mm continuum image with a factor of 2 higher resolution (in equatorial coordinates, black ellipse beam). The black contour shows the continuum at SNR=5. The magenta contours show the {\tt C4} ortho-H$_2$D$^+$ line integrated intensity at SNR=2, 3, 4, 5 (RMS$_{\rm mom0}$=10 mJy beam$^{-1}$ km s$^{-1}$, magenta ellipse beam). 
{\bf (d):} The new {\tt C7} 1.3 mm continuum image ({\tt C7Briggs0.5}, in equatorial coordinates, white ellipse beam). The field-of-view shows the two kernels in C1-Sa, which we name C1-Sa1 and C1-Sa2, both showing a centrally-peaked circular morphology. The yellow circles show the kernel definition (see \S\ref{subsec:frag}). The color bar is in Jy beam$^{-1}$.
\label{fig:showbinary}}
\end{figure*}

Figure \ref{fig:showbinary} shows the core fragmentation hierarchy.
Panel (a) shows an overview of the Dragon IRDC which is the dark 
absorption feature in the Spitzer $8\mu$m image from the 
GLIMPSE survey \citep[][]{2003PASP..115..953B,2009PASP..121..213C}.
The green contours show the 1.3 mm continuum mosaic from 
the {\tt C3} data. The C1-Sa and C1-Sb cores are in the Tail
of the Dragon 
\citep[inside the white dashed square, see][]{2019ApJ...873...31K}. 
This region has 
the highest column density of the IRDC \citep{2014ApJ...780L..29L}
and is dark at up to 100 $\mu$m \citep{2012A&A...547A..49R}.

Panel (b) shows the zoom-in view of the Tail region. 
Here we show the continuum emission,
along with the CO and SiO integrated intensities
in blue and orange contours, respectively.
The two lines trace molecular outflows from protostars. 
As will be shown in \S\ref{subsec:outflows},
there are three bipolar outflows in the Tail region.
Two of them (C1-Sa outflow 1 and C1-Sb outflow) were studied
in detail in T16 and K18 \citep[also see][]{2016ApJ...828..100F}.
The third one (C1-Sa outflow 2) is identified in this paper.
The arrows mark the three bipolar outflows. Here,
both lines are integrated between 74 and 80 km s$^{-1}$
to focus on C1-Sa outflow2. This velocity range  
encloses the system velocity of $\sim78$ km s$^{-1}$
\citep[][T16, K18]{1998ApJ...508..721C,2006A&A...450..569P}\footnote{For the velocities for C1-Sa outflow 1 and C1-Sb outflow, see the position-velocity diagrams in Figure 3 of T16.}.

Panel (c) shows the two continuum cores with a factor of 2
better resolution. Here, C1-Sa shows an elongated morphology
in the NE to SW direction. 
C1-Sb still appears to be a single weak structure. 
At a distance of 4.8 kpc, the two cores
have a projected separation of $\sim$0.067 pc (14,000 AU).
As shown in panel (b), C1-Sb coincides with the C1-Sa outflow2.
It is not clear if C1-Sb is
impacted by the outflow. But the outflow from C1-Sb
does not show any evidence of it being disrupted
(T16, K18). 

The magenta contours show the ortho-H$_2$D$^+$ integrated intensity.
The integration is between 78.8 and 80.0 km s$^{-1}$.
There is clear detection of ortho-H$_2$D$^+$.
The majority of the emission is 
between C1-Sa and C1-Sb, with some emission around C1-Sa.
The ortho-H$_2$D$^+$ spatial distribution is consistent with
the distribution of N$_2$D$^+$(3-2) from K18 (see their figure 3)
\footnote{Note that the velocity of ortho-H$_2$D$^+$
matches that of N$_2$D$^+$ (see K18) but has a slight
offset from the velocity defined by C$^{18}$O.}.
H$_2$D$^+$ is the first step of deuterium fractionation in 
cold starless cores
\citep[e.g.,][]{2013A&A...551A..38P,2015ApJ...804...98K}.
With its descendant N$_2$D$^+$, ortho-H$_2$D$^+$ has been 
used to search for the youngest dense cores 
\citep[e.g.,][]{2012ApJ...751..135P, 2020A&A...634A.115M,2020A&A...644A..34S,2021A&A...650A.202R}.
The presence of both ortho-H$_2$D$^+$ and N$_2$D$^+$ in the C1-S
region strongly indicates that the cores are in the earliest
evolutionary stage (K18)\footnote{The apparent offset between the core and the deuterated species is due to the heating from the protostellar core. Originally in \citet{2013ApJ...779...96T}, C1-S referred to the larger N$_2$D$^+$ core that enclosed C1-Sa and C1-Sb. With a higher angular resolution, K18 referred to C1-S as the N$_2$D$^+$ core in between the two protostellar cores. See K18 for more details.}. 

\begin{figure*}[htb!]
\centering
\epsscale{1.16}
\plotone{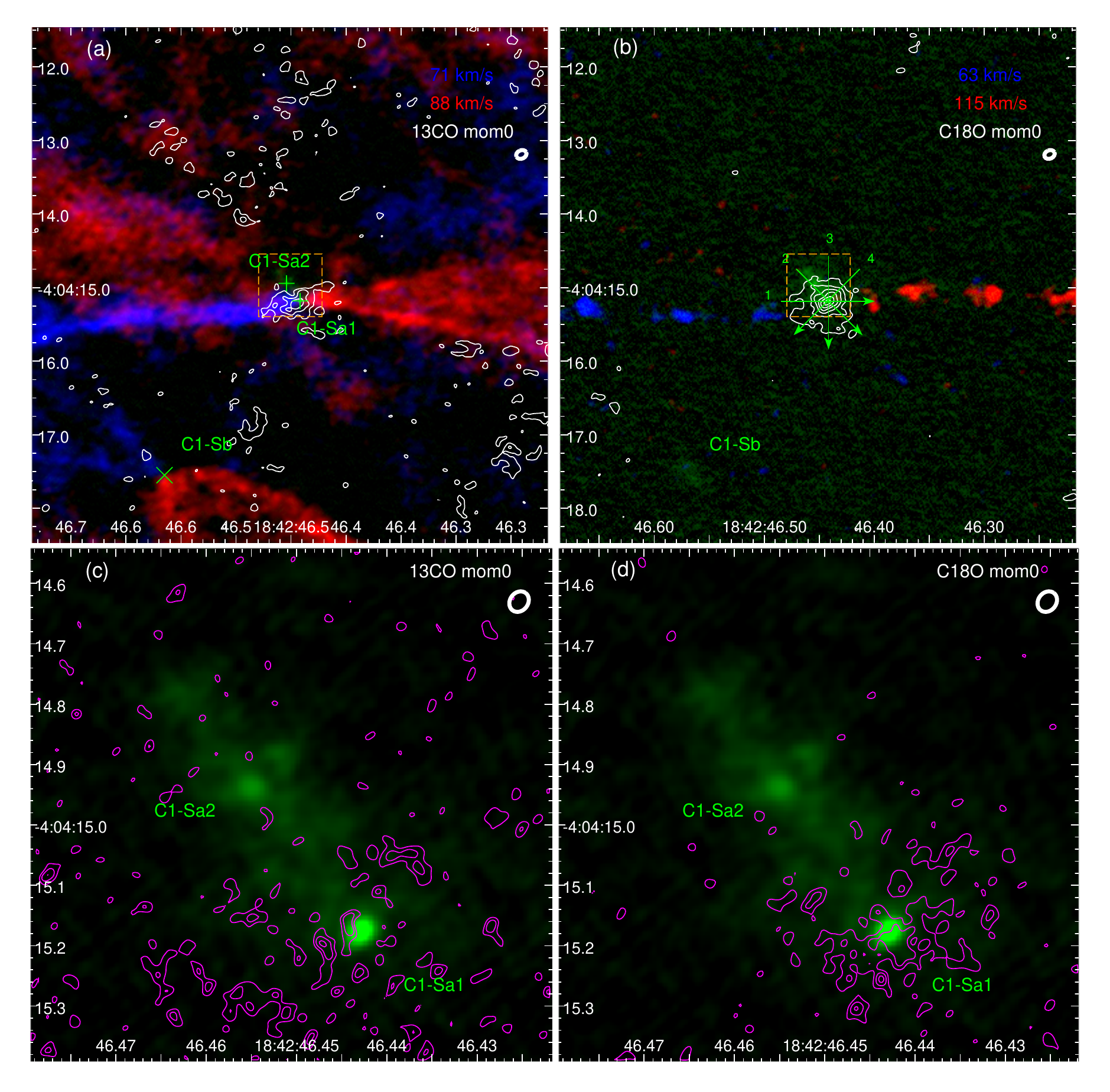}
\vspace{-10pt}
\caption{
{\bf (a):} RGB color map of C1-Sa and C1-Sa outflow1.
The green color shows the 1.3 mm continuum ({\tt C7Briggs0.5}),
ranging from 0 to 1 mJy beam$^{-1}$ (SNR=40) in linear scale. 
The blue color shows the CO channel at 71 km s$^{-1}$.
The red color shows the CO channel at 88 km s$^{-1}$.
The white contours show the $^{13}$CO integrated intensity 
({\tt C7Taper}) at SNR=3, 5, 7. The orange dashed rectangle shows
the field-of-view of panels (c)(d). {\bf (b):} 
The continuum (green) is the same as panel (a) but in square-root scale.
The blue color shows the SiO channel at 63 km s$^{-1}$.
The red color shows the SiO channel at 115 km s$^{-1}$.
The white contours show the C$^{18}$O integrated intensity
({\tt C7Taper}) at SNR=3, 5, ... 13.
The green arrows show positions of the cuts used for the position-velocity 
diagrams shown in Figure \ref{fig:pv}. Each cut is centered at C1-Sa1.
{\bf (c):} Comparison between 1.3 mm continuum (green, {\tt C7Briggs0.5})
and $^{13}$CO integrated intensity (magenta contour, {\tt C7Briggs0.5}).  
The contour levels are SNR=2, 3, 4.
{\bf (d):} Comparison between 1.3 mm continuum (green, {\tt C7Briggs0.5})
and C$^{18}$O integrated intensity (magenta contour, {\tt C7Briggs0.5}).  
The contour levels are SNR=2, 3, 4. Integration ranges and RMS noises
are presented in text in \S\ref{subsec:ld}.
\label{fig:showout}}
\end{figure*}

Panel (d) shows the zoom-in view of the C1-Sa core with the
high-resolution {\tt C7} data. 
The overall continuum emission follows the 
NE to SW elongation in panel (c).
The core fragments into two centrally-peaked, circular structures
(also see Figure \ref{fig:showout}(c)(d))
with several smaller and irregularly-shaped structures.
Hereafter, the circular structures are defined as kernels, 
and we name the two kernels  C1-Sa1 and C1-Sa2.
They have a projected separation of $\sim$1400 AU. 
The new {\tt C7} data confirms that C1-Sa is indeed
a binary in the making\footnote{The continuum emission immediately to 
the north-west of C1-Sa2 shows a hint of a centrally-peaked morphology.
This structure could eventually develop into a third kernel.}.

We carry out 2D Gaussian fitting to the kernels. The
C1-Sa2 fitting is affected by its surrounding emission.
So we simply adopt the centers from the fitting but
define the kernels with circles. The yellow circles in
Figure \ref{fig:showbinary}(d) show the kernel definition.
For C1-Sa1 the circle is at 3$\sigma$ level of the Gaussian fitting.
For C1-Sa2 the circle is smaller than the 3$\sigma$ level
of the fitting but covers the circular kernel visually.
The position error for C1-Sa1 fitting is 3\% of a pixel (0.003\arcsec);
for C1-Sa2 is 10\% of a pixel.
Table \ref{tab:kernel} summarizes the kernel definition.
From now on, we assume a spherical geometry in 3D 
for the two kernels.

\begin{deluxetable*}{cccccccccc}
\tabletypesize{\footnotesize}
\tablecaption{Kernel Definition and Properties \label{tab:kernel}}
\tablehead{
\colhead{Kernel} &
\colhead{RA} &
\colhead{DEC} &
\colhead{$R$} &
\colhead{$D$} &
\colhead{$I_{\rm peak}$} &
\colhead{$S_{\rm 1.3}$} &
\colhead{$M$} &  
\colhead{$V_{\rm LSR}$} &
\colhead{$\Delta V$}\\
\colhead{} &
\colhead{J2000} & 
\colhead{J2000} &
\colhead{arcsec} &
\colhead{AU} &
\colhead{mJy beam$^{-1}$} &
\colhead{mJy} &
\colhead{M$_\odot$} &
\colhead{km s$^{-1}$} &
\colhead{km s$^{-1}$}
}
\startdata
C1-Sa1 & 18:42:46.4429 & -4:04:15.175 & 0.044 & 420 & 1.46(0.02) & 3.4(0.6) & 0.55(0.09) & 78.59(0.06) & 0.60(0.06) \\
C1-Sa2 & 18:42:46.4550 & -4:04:14.939 & 0.044 & 420 & 0.54(0.02) & 2.2(0.6) & 1.8(0.5) & - & -
\enddata
\tablecomments{The kernel radii are not forced to be the same in the fitting.}
\end{deluxetable*}

\subsection{C1-Sa Outflows}\label{subsec:outflows}

In Figure \ref{fig:showbinary}(b),
there are two bipolar outflows that are visually
associated with C1-Sa. The first (C1-Sa outflow1) was identified
in T16 and K18. The second (C1-Sa outflow2) is a bit ambiguous 
as its two lobes  show up in the same CO channel maps  (T16).
This is unlike most other symmetric outflows that typically show one lobe
in blue-shifted channels and the other in red-shifted channels,
or monopolar outflows \citep[e.g.,][]{2013ApJ...778...72F}.
However, the SiO(5-4) integrated intensity contours nicely
overlap with the CO counterparts in panel (b), especially in
the north-west end (the bifurcation feature).
The morphological match between the two outflow
tracers, which is seen in many other outflows in this IRDC 
\citep[see][]{2019ApJ...874..104K}, strongly indicates that C1-Sa 
outflow2 is indeed a bipolar outflow. The axis of this outflow 
is probably parallel to the plane of the sky, which is why the 
two lobes always appear in same channels. 
\citet{2016ApJ...828..100F} saw the same NW to SE
structure in SiO(2-1) at a lower resolution. They suspected that
the bipolar SiO(2-1) emission belonged to a second outflow, consistent
with our findings here.
Visually, C1-Sa outflow2 clearly points toward C1-Sa,
which is why this outflow is named after C1-Sa.
However, C1-Sa outflow2 is not like C1-Sa
outflow1 that connects to the C1-Sa core. There is a small
gap between the outflow and the core. 

Figure \ref{fig:showout}(a) shows the C1-Sa outflow1 in CO with 
the latest {\tt C7} data. The outflow is clearly launched
from C1-Sa1, the brightest kernel of the two.
The red lobe shows a nice fan-like morphology, while the blue
lobe is more like a narrow jet. The C1-Sb bipolar outflow is
also visible to the south-east. Notably, C1-Sb is relatively
fuzzy in the high-resolution continuum data
(see panel (b), for comparison see K18 Figure 1),
suggesting that the core is not very centrally-peaked. However, 
it launches a bipolar outflow. But the outflow is not as collimated
as those from C1-Sa.

Figure \ref{fig:showout}(b) shows C1-Sa outflow1 in SiO.
Here we see multiple knots in the outflow, probably indicating
accretion bursts in C1-Sa1 
\citep[see another example in][]{2015Natur.527...70P}. 
A visual inspection of the larger area shows that there are 
at least 6 burst knots in each lobe of the outflow.
The knots are approximately equally-spaced, with
a typical projected separation $l_\perp$=1\arcsec,
equivalent to 4800 AU or 0.023 pc in the plane of the sky.
The line-of-sight velocity $v_\parallel$ for the blue knots
is 15 km s$^{-1}$ and for the red is 37 km s$^{-1}$ (adopting
a system velocity of 78 km s$^{-1}$)\footnote{The blue knots are clearly seen between 40-70 km s$^{-1}$. The red knots are between 100-125 km s$^{-1}$. So the fastest knot velocity is about 40 km s$^{-1}$.}, which implies more than a
factor of two difference between the red and blue (also seen in T16).
Assuming the same outflow velocity between the two lobes,
the mismatch $v_\parallel$ indicates that the blue lobe has
a larger $i$. Alternatively, the blue lobe is plunging into
denser gas so that it is decelerated. 

If we simply adopt a line-of-sight knot
velocity of 30 km s$^{-1}$, the burst interval 
$\delta t=l_\perp/v_\parallel/tan(i)\approx450$ yr,
where the inclination angle relative to our line-of-sight
$i$ is set at 60\arcdeg\ to be consistent with T16\footnote{Note T16 used a mass-weighted outflow velocity while we are using the channel velocity for the knots. The dynamical timescale for C1-Sa outflow1 was estimated to be $\sim10$ kyr}.
The very rough estimation
suggests that the accretion bursts take place regularly on a timescale on the order of 500 yr. 

\subsection{Line Detection}\label{subsec:ld}

\begin{figure*}[htb!]
\centering
\epsscale{1.15}
\plottwo{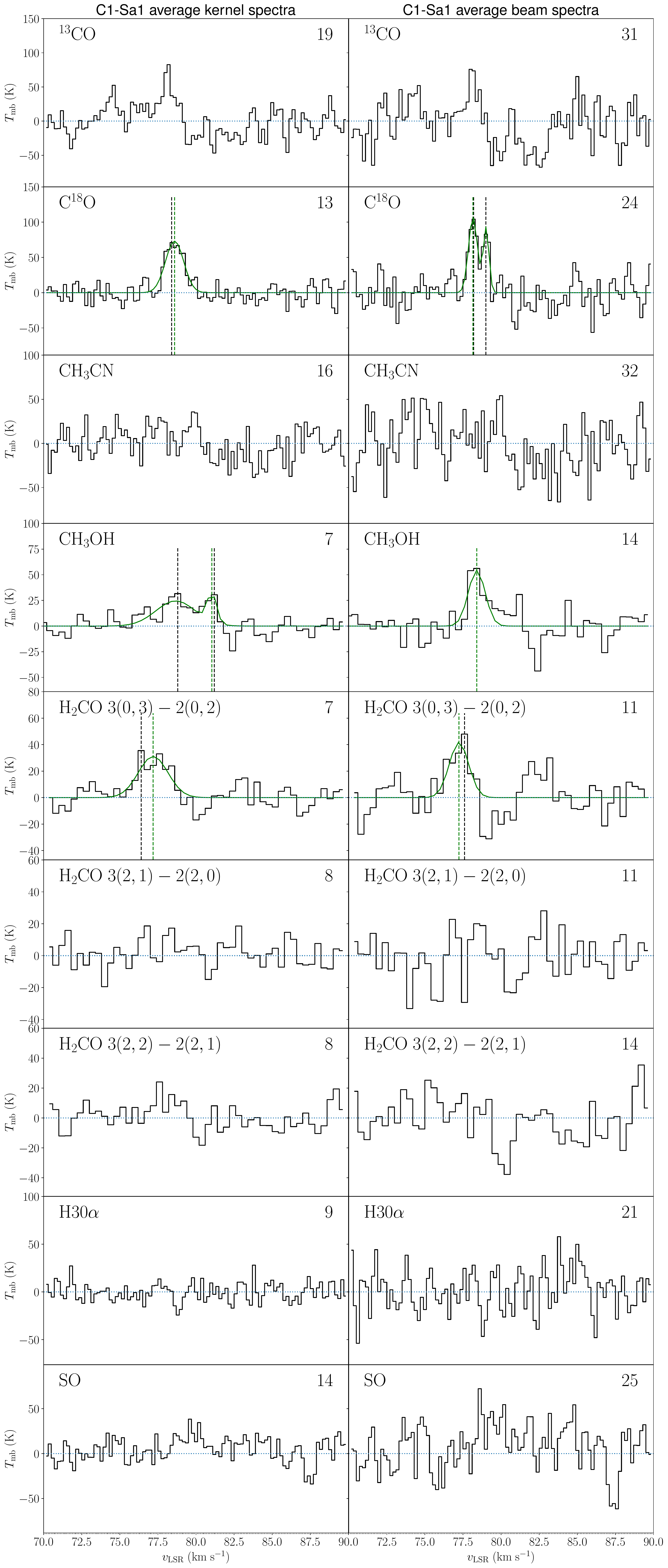}{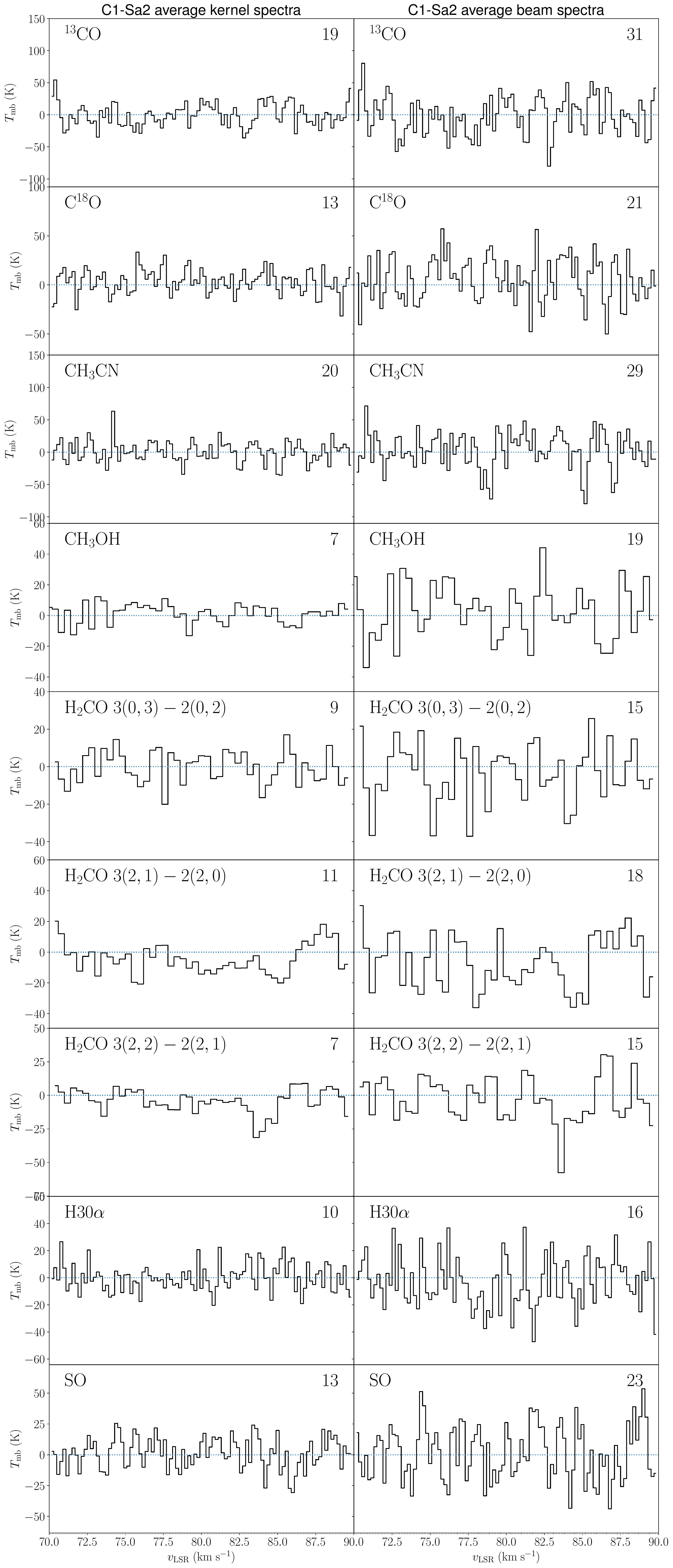}
\caption{
C1-Sa1 spectra (left) and C1-Sa2 spectra (right) from {\tt C7Briggs0.5}. For each source, we show both average spectra within the kernel and the beam. If a local peak (black dashed vertical line) has at least 3 adjacent channels with SNR$\geq$2, it is identified as a component and fitted with a Gaussian (green curve). The line noise RMS$_{\rm line}$ in K is shown at the top-right. The black vertical dashed line indicates the line peak. The green vertical dashed line marks the Gaussian peak.
\label{fig:spec}}
\end{figure*}

Figure \ref{fig:showout}(a)(b) show the integrated intensity
contours of $^{13}$CO and C$^{18}$O ({\tt C7Taper}).
The integration is over 77.3 km s$^{-1}$ to 80.0 km s$^{-1}$,
with RMS$_{\rm mom0}$ $\sim$2.2 mJy beam$^{-1}$ km s$^{-1}$.
One can see that both molecular lines clearly 
show a centrally-peaked emission structure at C1-Sa1. 
In the $^{13}$CO contour map, there is a concave emission
pattern around C1-Sa2. In C$^{18}$O, C1-Sa2 is on the 
periphery of the contours. C1-Sa2 does not show a centrally-peaked 
molecular line emission. 

In Figure \ref{fig:showout}(c)(d), we zoom in to show the 
detailed emission around the two kernels. Here we overlay
the {\tt C7Briggs0.5} $^{13}$CO and C$^{18}$O integrated 
intensity contours on top of the {\tt C7} continuum image.
The integration is over 77.3 km s$^{-1}$ to 80.0 km s$^{-1}$,
with RMS$_{\rm mom0}$ $\sim$1.8 mJy beam$^{-1}$ km s$^{-1}$.
Again, C1-Sa2 is neither detected in $^{13}$CO 
nor C$^{18}$O, while C1-Sa1 shows clumpy detection in both
lines. There is a void of C$^{18}$O toward C1-Sa2,
indicating that the NE part of
the dust elongation is  cold enough  that the C$^{18}$O is 
reduced from gas-phase due to freeze-out,
which is consistent with the 
spatial distribution of ortho-H$_2$D$^+$ 
(Figure \ref{fig:showbinary}(c)). C1-Sa2 is either 
starless or at a very early evolutionary stage
with undetectable lines or outflows (see below).

Comparing C1-Sa1 and C1-Sa2, we argue that C1-Sa outflow2
could not have been launched by C1-Sa2.
C1-Sa outflow2 is much more extended than C1-Sa outflow1
in the sky plane. If it was indeed from C1-Sa2,
C1-Sa2 should at least show similar $^{13}$CO and C$^{18}$O
detection as C1-Sa1 because of heating of the environment by the protostar.
One could argue that C1-Sa outflow2
from C1-Sa2 was significantly reduced for some reason a while ago. 
The gap between C1-Sa and one lobe from C1-Sa
outflow2 is about 1\arcsec, corresponding to 4800 AU which
takes 770 yr for a 30 km s$^{-1}$ outflowing gas to traverse. 
Given the length of the C1-Sa outflow2, the presumed 
C1-Sa2 protostellar accretion must have lasted around
ten thousand years (one outflow lobe is about 20\arcsec,
and assuming a velocity of 30 km s$^{-1}$, the timescale 
is 16000 yr). It is hard to imagine that C1-Sa2 had a
protostar for about ten thousand years then significantly
reduced accretion and resumed CO freeze-out within a thousand years
(the timescale of the gap). 
In addition, we see no hint of outflowing gas connecting to C1-Sa2
(see \S\ref{subsec:out2}).
A more plausible scenario is that C1-Sa2 is still starless.

In Figure \ref{fig:spec}, we plot molecular line spectra
for the two kernels. For each kernel, we show 
both a spectrum averaged over the size of the kernel and 
spectrum average over the size of the beam,
centered at the kernel position. Note that the kernel size
($\sim$0.09\arcsec) is $\sim$3 times the beam size
($\sim$0.03\arcsec). We  adopt an automated 
routine to find and fit Gaussian profiles in the spectra. Only
those velocity components with at least 3 adjacent channels
greater than twice the spectral noise are considered.

C1-Sa1 shows clear detection in 
C$^{18}$O, CH$_3$OH 4(2)-3(1) E1 vt=0, and H$_2$CO 3(0,3)-2(0,2).
The $^{13}$CO
detection is marginal and not captured by the auto-fitting
routine. A visual inspection of the channel maps shows that
the C1-Sa outflow1 is detected in CH$_3$OH 4(2)-3(1) E1 vt=0, 
H$_2$CO 3(0,3)-2(0,2),
H$_2$CO 3(2,1)-2(2,0), H$_2$CO 3(2,2)-2(2,1), SO, and SiO. So the
detection of CH$_3$OH and H$_2$CO in the C1-Sa1
spectra is probably contaminated by the outflow. 
C1-Sa2 shows no clear detection in any of the lines, 
which is consistent with it being starless. 

\begin{figure*}[htb!]
\centering
\epsscale{1.15}
\plotone{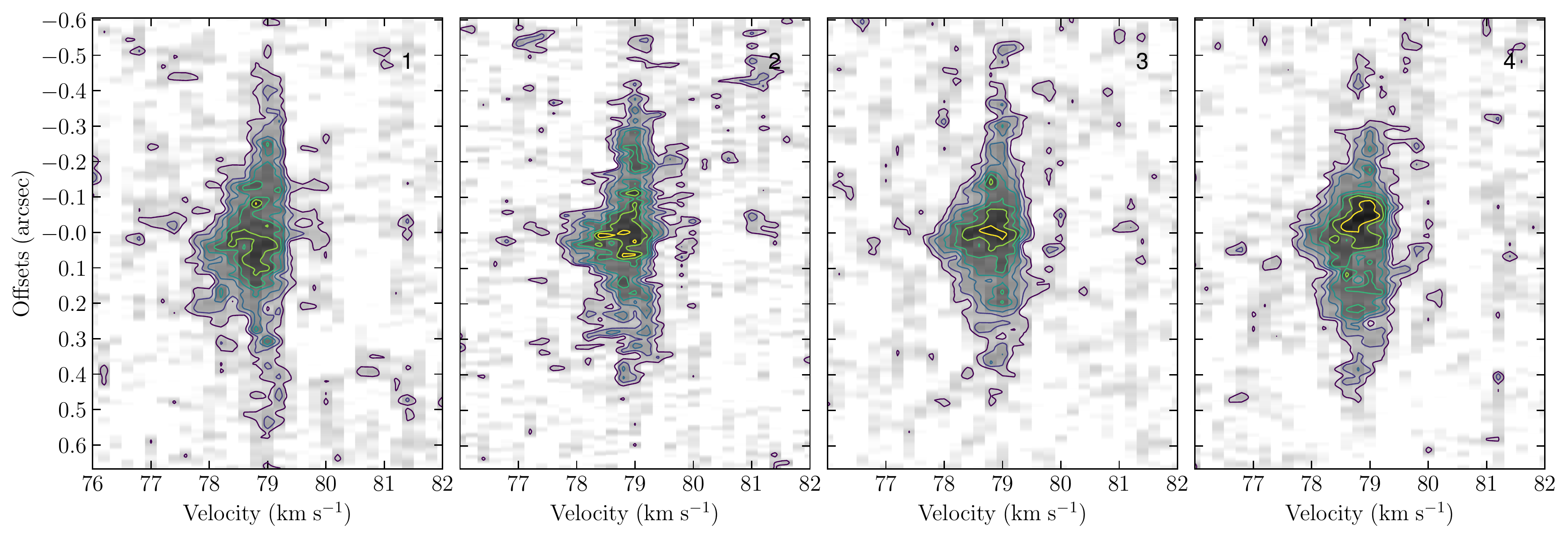}
\caption{ C$^{18}$O(2-1)
position-velocity diagrams across C1-Sa1 (using {\tt C7Briggs1}). Each panel 
shows the diagram along the  cut as denoted by the relevant
arrow in Figure \ref{fig:showout}(b). The offset direction 
follows the arrow direction. C1-Sa1 is at zero offset. The
color range is from 0 to 0.01 Jy beam$^{-1}$ (SNR=10). 
The color map is also shown in contours, with
levels ranging from SNR=2 to SNR=10 at a step of SNR=1.
\label{fig:pv}}
\end{figure*}

In C1-Sa1, the C$^{18}$O kernel spectrum shows a Gaussian 
profile. The fitted line centroid velocity is 
$78.59\pm0.06$ km s$^{-1}$, 
with a velocity dispersion of $0.60\pm0.06$ km s$^{-1}$. 
However, the beam spectrum shows a double-peak feature
with the red peak being weaker. Their separation is 
0.87$\pm$0.06 km s$^{-1}$. The double-peak feature
only appears at the kernel center. Figure \ref{fig:pv}
shows C$^{18}$O position-velocity diagrams along four cuts through C1-Sa1
as shown in Figure \ref{fig:showout}(b). They all show
a line-broadening near offset 0 (the location of the kernel),
which is consistent with the spectra. Note, the line-broadening
is most obvious at blue-shifted velocities but suppressed at
red-shifted velocities in the position-velocity diagram. As will be shown
in \S\ref{subsec:phys}, the C$^{18}$O line is optically thick.
Therefore, the double-peak feature is probably due to
self-absorption by the red-shifted colder gas on the near side.
Gravitational infall at the kernel center can explain these spectral
features \citep[e.g., the blue profile in][]{1999ARA&A..37..311E}. 
On the other hand, there is no clear velocity
gradient across C1-Sa1 in any of the four position-velocity diagrams.
They all show a main velocity component at $\sim$79 km s$^{-1}$.
The lack of gradient indicates a lack of rotation across
C1-Sa1 (to the accuracy of the velocity resolution of
$\sim0.2$ km s$^{-1}$) above the scale of $\sim$300 AU 
({\tt C7Briggs1} beam size of 0.06\arcsec). 

\subsection{Physical Properties}\label{subsec:phys}

In C1-Sa, K18 adopted a fiducial dust temperature of 20 K
for the entire core and derived a mass of 33.5 M$_\odot$.
Hereafter, we adopt their fiducial temperature and mass for C1-Sa.
The corresponding column density is 22000 M$_\odot$ pc$^{-2}$
($\Sigma=4.7$ g cm$^{-2}$ or $N_H=2.0\times10^{24}$ cm$^{-2}$).
For the moderately coagulated thin ice mantle model
$\kappa_\nu=0.899$ cm$^2$ g$^{-1}$ at 1.3 mm
\citep{1994A&A...291..943O}, the core optical depth $\tau_{1.3}=\Sigma\kappa_\nu/R_{\rm gd}=0.028$
\citep[assuming a gas-to-dust mass ratio $R_{\rm gd}=141$,][]{2011piim.book.....D},
i.e., still optically thin\footnote{If the C1-Sa temperature were 15 K,
the core mass would be a factor of 1.5 higher (K18) and the core would
still be optically thin at 1.3 mm.}. The  particle number density 
is $n=1.3\times10^7$ cm$^{-3}$ (assuming a mean molecular weight 
per free particle of 2.37, K18). At such high density,
the gas temperature is expected to be the same as the dust temperature (20 K),
and thus the sound speed, 
$c_s$, should be about 0.3 km s$^{-1}$.
With these properties, the Jeans length is about $0.005$ pc (1100 AU),
which is comparable
to the projected separation between C1-Sa1 and C1-Sa2
(\S\ref{subsec:frag}). Unless the elongated structure of C1-Sa 
is significantly inclined with respect to the plane of the sky,
in which case the 3D separation is much larger than the projected separation,
the distance between the kernels suggests a thermal
fragmentation scenario for the binary formation.
At present no observations clearly constrain the inclination.

We fit the C1-Sa core C$^{18}$O(2-1) spectra from the
{\tt C2} data (K18) and {\tt C3} data (K21)
at the scale of 1.8\arcsec (averaged within a diameter of
1.8\arcsec, or 0.042 pc, which was the diameter of C1-Sa in 
the K18 definition). Both fits
give a velocity dispersion $\sigma_{18}=0.30\pm0.02$ km s$^{-1}$
which is approximately the sound speed $\sigma_{18}\approx c_s$. 
This means the non-thermal motion is sub/transonic as we subtract
the thermal component from the C$^{18}$O line dispersion.
If the non-thermal component is tracing turbulence,
then the data shows that turbulence becomes less dominant
at the scale of C1-Sa, which is consistent with the thermal
fragmentation argument above. 

Interestingly, if we fit the spectra
at a scale of 1.0\arcsec\ (average spectrum over the size of the scale), 
the {\tt C2} data gives $\sigma_{18}=0.33\pm0.01$ km s$^{-1}$
while the {\tt C3} data gives $\sigma_{18}=0.35\pm0.03$ km s$^{-1}$. 
If we fit the spectra at a scale of 0.5\arcsec,
the {\tt C2} data gives $\sigma_{18}=0.34\pm0.01$ km s$^{-1}$
while the {\tt C3} data gives $\sigma_{18}=0.36\pm0.03$
km s$^{-1}$. The fact that the line dispersion is constant/increases 
at smaller scales (from 1.8\arcsec\ to 0.5\arcsec) indicates 
that it is not tracing turbulence decay from larger scales, 
consistent with the above result that turbulence is less important.
At an even smaller scale (the kernel scale of C1-Sa1 0.088\arcsec),
the line dispersion becomes $0.60\pm0.06$ km s$^{-1}$ (\S\ref{subsec:ld}).
This increasing linewidth could trace the gravitational collapse
that accelerates closer to C1-Sa1 
\citep[e.g., see][]{2011MNRAS.411...65B}. Or, it is due to the
line-broadening from high optical depth (see below). Another possibility
is that it is tracing the rotation inside C1-Sa1 (\S\ref{subsec:out2}).

We compute the kernel mass with the 
following equation
\begin{equation}\label{equ:mass}
M=\frac{S_\nu D^2 R_{\rm gd}}{B_\nu (T_{\rm dust}) \kappa_\nu},
\end{equation}
where $S_\nu$ is the continuum flux density,
$D$ is the distance (4.8 kpc),
$R_{\rm gd}$ is the gas-to-dust mass ratio
\citep[141,][]{2011piim.book.....D}, $B_\nu (T_{\rm dust})$
is the Planck function at the dust temperature, and $\kappa_\nu$
is the dust opacity. We adopt $\kappa_\nu=0.899$
cm$^2$ g$^{-1}$ for the moderately coagulated thin ice mantle model 
from \citet{1994A&A...291..943O}. If we again adopt a dust 
temperature of 20 K, the mass for C1-Sa1 is 2.6 M$_\odot$ 
(but see below), and for C1-Sa2 is 1.6 M$_\odot$. 

The above C1-Sa1 mass at 20 K corresponds to a volume density of
$n_{\rm H_2}=8.2\times10^9$ cm$^{-3}$, and a column density of
$N_{\rm H_2}=3.5\times10^{25}$ cm$^{-2}$. Assuming no CO freeze-out
in C1-Sa1, the corresponding C$^{18}$O column density is 
$N_{\rm C^{18}O}=1.9\times10^{19}$ cm$^{-2}$, assuming the 
ISM abundance [C$^{18}$O/H$_2$]=$5.4\times10^{-7}$
\citep{1994ARA&A..32..191W,2015A&A...573A..30G}.
The C1-Sa1 C$^{18}$O
kernel spectrum had a line dispersion of 0.60 km s$^{-1}$, corresponding 
to a linewidth of 1.4 km s$^{-1}$. Putting these numbers into the 
\href{https://home.strw.leidenuniv.nl/~moldata/radex.html}{RADEX}
code \citep{2007A&A...468..627V},
assuming a gas kinetic temperature of 20 K, the C$^{18}$O
line optical depth $\tau_{18}$ reaches 1500. 

However, under such conditions,
the line intensity is only 15 K based on RADEX,
much smaller than what we observe (see Figure \ref{fig:spec}).
To crank up the line intensity, one has to increase the kinetic
temperature which is the only uncertain variable in the above
calculation. In short, if C1-Sa1 has a kinetic temperature of
75 K, the line intensity estimated by RADEX of 70 K would match the 
C1-Sa1 kernel spectrum in Figure \ref{fig:spec},
and the corresponding mass would be 0.55 M$_\odot$ 
(still using Eq. (\ref{equ:mass}) but $T_{\rm dust}=75$ K).
Note, the Jeans mass in C1-Sa
(the enclosing core) is 0.087 M$_\odot$, which is significantly
smaller than the kernel mass. Perhaps magnetic fields provide additional 
support. For instance, \citet{2020ApJ...895..142L} showed an ordered
magnetic field which is well aligned with the elongation of C1-Sa
(NE-SW). Alternatively, they probably have accreted material from
C1-Sa, i.e., their larger mass is not the mass that directly results
from the fragmentation.

The C1-Sa1 mass is a factor of 3 smaller than C1-Sa2.
The difference can be explained by the protostellar accretion in C1-Sa1.
For instance, if the two kernels had a similar initial mass, 
then the missing mass in C1-Sa1 could have been accreted by the 
embedded protostar. With the new estimation for C1-Sa1
(0.55 M$_\odot$), the volume density becomes
$n_{\rm H_2}=1.8\times10^9$ cm$^{-3}$, and the column density becomes
$N_{\rm H_2}=7.4\times10^{24}$ cm$^{-2}$. Assuming the same
ISM abundance [C$^{18}$O/H$_2$]=$5.4\times10^{-7}$,
the C$^{18}$O column density becomes
$N_{\rm C^{18}O}=4.0\times10^{18}$ cm$^{-2}$. The updated RADEX
calculation gives an optical depth of 35, i.e.,
the C$^{18}$O line is optically thick.

For C1-Sa2, using the mass calculation at 20 K,
the volume density is 
$n_{\rm H_2}=5.3\times10^9$ cm$^{-3}$, 
and the column density is
$N_{\rm H_2}=2.2\times10^{25}$ cm$^{-2}$.
If there is no CO freeze-out in C1-Sa2,
the corresponding C$^{18}$O column density is 
$N_{\rm C^{18}O}=1.2\times10^{19}$ cm$^{-2}$.
Assuming a linewidth of $\sim$1.4 km s$^{-1}$,
the C$^{18}$O(2-1) line intensity based on RADEX
is 15 K, which is the same as C1-Sa1 at 20 K
because the line emission is saturated with 
an optical depth of 970. The 15 K line intensity
is non-detectable in the kernel spectrum in Figure
\ref{fig:spec} which has an RMS$_{\rm line}$ noise of 12 K.
Only if we increase the kernel temperature to 40 K
would the C$^{18}$O(2-1) line intensity be high enough to be
detectable with SNR of 3, albeit still being optically thick.
Therefore, the non-detection of C$^{18}$O in C1-Sa2
sets an upper limit of the kernel temperature to 40 K.

The surface density of C1-Sa1 is 
$\Sigma=34$ g cm$^{-2}$. The corresponding optical depth is 0.20,
i.e., optically thin. Meanwhile, we estimate the surface density
for C1-Sa2 to be $\Sigma=100$ g cm$^{-2}$. The optical depth is 0.62.
The continuum flux for C1-Sa1 is 3.4 mJy, for C1-Sa2 is 2.2 mJy.
The corresponding Planck brightness temperature (the equivalent temperature
with which the Planck function gives the flux) for C1-Sa1 is 15 K, 
for C1-Sa2 is 11 K. Since the dust continuum is optically thin,
the above brightness temperatures
are lower limits of the kernel temperature.

\section{Discussion}\label{sec:disc}

\subsection{C1-Sa Outflow2 Origin from C1-Sa1}\label{subsec:out2}

\begin{figure*}[htb!]
\centering
\epsscale{1.15}
\plotone{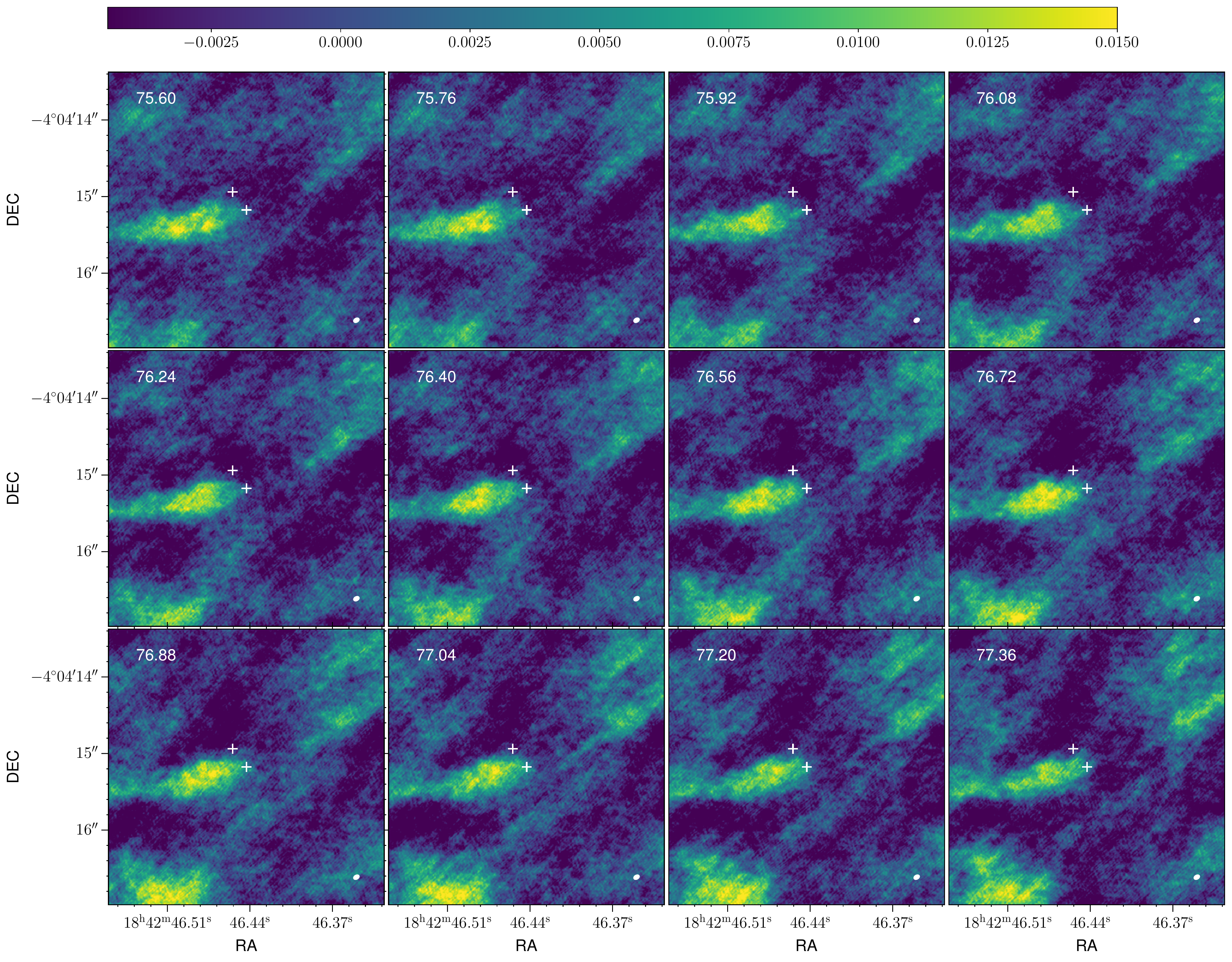}
\caption{
Zoom-in CO channel maps at C1-Sa1 outflow2 velocities ({\tt C7Natural}). The channel velocity is labeled at the top-left in km s$^{-1}$. The linear-scale color range is between SNR=-3 and SNR=10 (RMS$_{\rm line}$=1.5 mJy beam$^{-1}$). The two kernels are marked as white crosses (C1-Sa1 at the map center). C1-Sa outflow1 is the bright structure in the middle pointing eastward. Only its blue lobe is visible (see Figure \ref{fig:showout}). C1-Sa outflow2 is the bright structure from lower-left to upper-right.
\label{fig:chanco}}
\end{figure*}

In \S\ref{subsec:outflows}, we have shown that C1-Sa launches two
outflows, include C1-Sa outflow2. In \S\ref{subsec:ld},
we have argued that C1-Sa outflow2 is not from C1-Sa2.
Therefore, C1-Sa1 is the only candidate for the launching source
of C1-Sa outflow2. Figure \ref{fig:chanco} 
shows the spatial relation between the two kernels and the outflows.
Here, we show CO channel maps from {\tt C7Natural}, 
focusing on channels in which the C1-Sa outflow2 is
present. Again, C1-Sa outflow1 is clearly from C1-Sa1.
Meanwhile, C1-Sa outflow2, though not clearly contacting to the
two kernels, appears to be spatially closer to C1-Sa1.
There is some weak emission feature (SNR$\geq$2) extending from 
C1-Sa outflow2 and pointing toward C1-Sa1.  Here we argue that C1-Sa 
outflow2 originates from the C1-Sa1 kernel.

Two scenarios could explain the origin of C1-Sa outflow2.
In the first scenario, there is only one protostar in C1-Sa1.
It initially launched C1-Sa outflow2.
About 770 yr ago (the gap between C1-Sa outflow2 and C1-Sa1, 
see \S\ref{subsec:ld}), the accretion was significantly reduced
and a new outflow (C1-Sa outflow1) was launched at a different
direction. In this scenario, the change in outflow direction 
could have been caused by a change in direction of the 
accretion disk angular momentum vector. However, based on the
outflow bursts, the age of C1-Sa outflow1 should be at least 6 times
the burst interval (6$\times$450 yr=2.7 kyr, see \S\ref{subsec:outflows}). 
So C1-Sa outflow1 already existed before C1-Sa outflow2 weakened.

A more plausible scenario is that C1-Sa1 hosts a tight binary
that is not resolved by the ALMA synthesized beam of 140 AU. One 
launched C1-Sa outflow2 first and the other launched C1-Sa outflow1 
later. Both outflows co-existed for a while and one of them 
significantly weakened 770 yr ago. 
The double-peak C$^{18}$O line (optically thick)
could show the overall infall in C1-Sa1 
\citep[the ``blue profile'',][]{1999ARA&A..37..311E},
or the radial velocity of the tight binary
(each peak tracing a member), 
or the rotational velocity of an edge-on disk
(a circumbinary disk or the circumstellar disks)
that is not resolved. 

\subsection{Further Fragmentation}\label{subsec:morefrag}

\begin{figure*}[htb!]
\centering
\epsscale{1.15}
\plotone{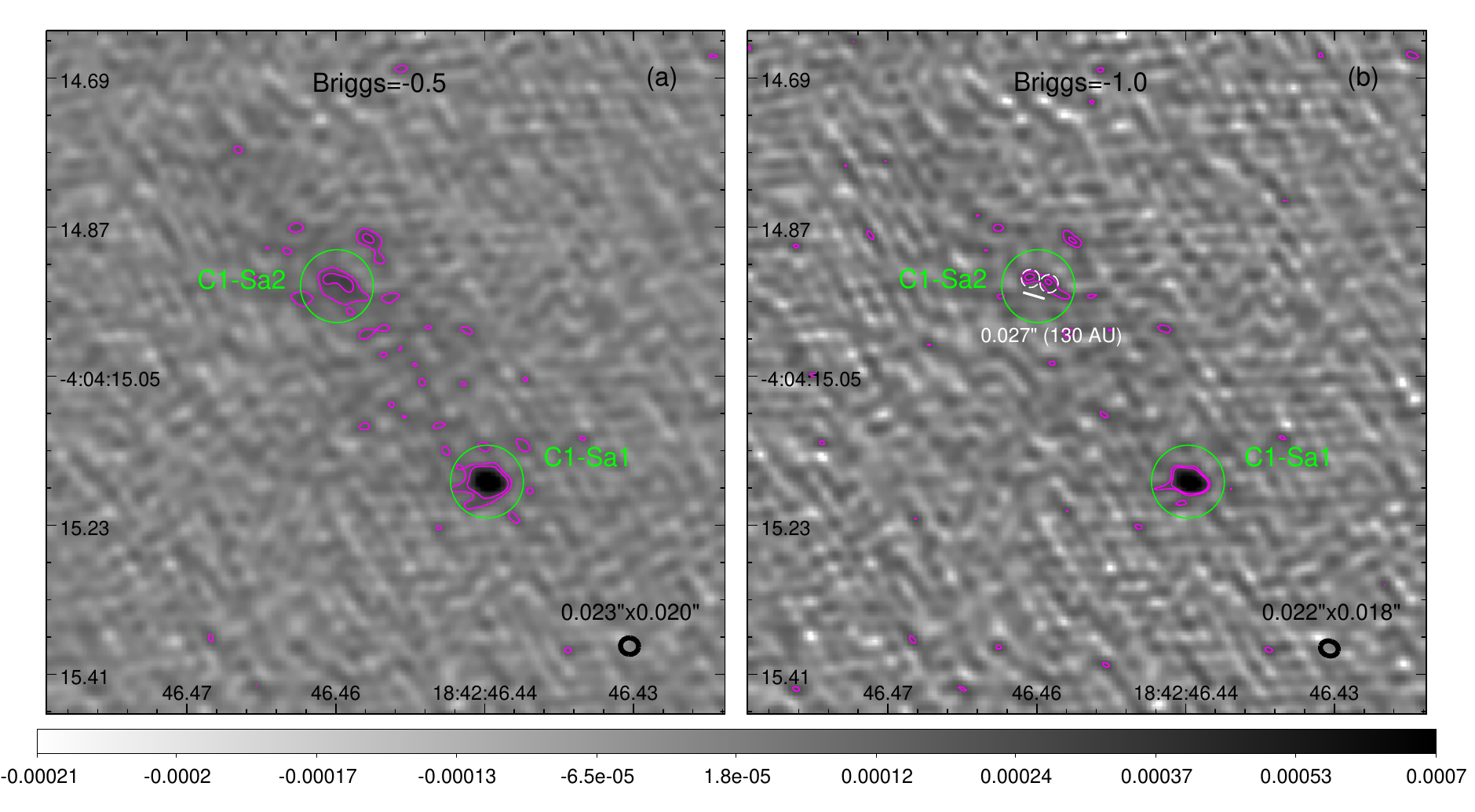}
\caption{
{\bf (a):} {\tt C7Briggs-0.5} 1.3 mm continuum image. 
The linear color scale is in unit of Jy beam$^{-1}$.
The magenta contours show emission at SNR=3,5 with RMS$_{\rm cont}$=50$\mu$Jy beam$^{-1}$.
{\bf (b):} Same field of view and same color scale as panel (a),
but with {\tt C7Briggs-1} 1.3 mm continuum.
The magenta contours show emission at SNR=3,4 with RMS$_{\rm cont}$=70$\mu$Jy beam$^{-1}$.
The white dashed circles show the definitions of the sub-kernels
of the potential pre-binary, both having a diameter of 0.022\arcsec,
or 100 AU. Their separation (0.027\arcsec\ or 130 AU)
is marked by the white segment.
\label{fig:mf}}
\end{figure*}

In Figure \ref{fig:mf}, we show the {\tt C7} 
1.3 mm continuum image with an even higher resolution 
than the previous images. Panel (a) shows the {\tt C7Briggs-0.5} 
image with a synthesized beam size of about 0.022\arcsec. 
Panel (b) shows the {\tt C7Briggs-1} image with 
a beam size of approximately 0.02\arcsec\ (96 AU).
With the higher resolution, we see C1-Sa1 starts to show irregular shapes
at its periphery, which seems to support the tight binary scenario
because a single protostar should maintain a form of symmetry
in its morphology\footnote{We caution that the irregular shape
could be the result of interferometric artifacts, which can be
improved with more uv-coverage.}.
Still, no clear signature of disks is obvious in the image,
constraining the disk sizes to be smaller than 96 AU.
Higher angular resolution observations are needed to confirm 
the potential tight binary.

In Figure \ref{fig:mf}, C1-Sa2 splits into two sub-kernels,
or a potential pre-binary\footnote{A binary that is in prestellar phase, i.e., no protostars yet but self-gravitated.}, both detected at SNR=4.
Their projected separation is $\sim$130 AU.
In panel (b), we simply define the sub-kernels with two 
circles that cover the continuum emission (white dashed circles).
They both have a flux density of 290$\pm70\mu$Jy (due to rounding),
translating to a mass of $\sim0.17\pm0.04$ M$_\odot$ assuming a 20 K 
dust temperature (Eq. (\ref{equ:mass})).
The free particle number density of 
C1-Sa2 is $n=6.1\times10^9$ cm$^{-3}$.
The corresponding Jeans scale is 57 AU, and
the corresponding Jeans mass is 0.0040 M$_\odot$.
In \S\ref{subsec:phys}, we set a
temperature upper limit of 40 K. Since C1-Sa2 has no 
outflow detection and there is ortho-H$_2$D$^+$ (Figure
\ref{fig:showbinary}) and N$_2$D$^+$ (K18) in the vicinity,
its temperature cannot be much higher than 20 K
based on deuterium chemistry \citep[e.g.,][]{2015ApJ...804...98K}.

Note in {\tt C7Briggs0.5} (Figure \ref{fig:showout}(b)),
C1-Sb is not as centrally-concentrated as C1-Sa2
and has a factor of
$\ga$3 weaker 1.3 mm emission than C1-Sa2, yet its outflow is 
clearly detected. So either C1-Sa2 has no outflow or its
outflow is still too weak/small to be detected, putting it
at a very early stage with likely a lower temperature. The blob
to the north-west (\S\ref{subsec:frag}) remains singular.
Altogether, the whole C1-Sa core is potentially
developing a quintuple system within a scale of $\sim$1000 AU.
However, we caution that the fragmentation of C1-Sa2 is 
to be further confirmed with higher sensitivity observations.

\subsection{Binary Star Formation}\label{subsec:bsf}

The massive ($\sim30$ M$_\odot$), 100 $\mu$m-dark core C1-Sa 
shows a hierarchical structure. The core itself harbors a 
binary system at the two ends of an elongated structure.
Each member of the binary shows signs of further fragmentation.

If the misaligned outflows in C1-Sa1 indeed indicate a tight
binary with misaligned disks, then the misalignment
argues against disk fragmentation.
On the other hand, turbulence is sub/transonic at the sub-core 
scale based on the C$^{18}$O line fitting. So  the 
origin of the hypothesized binary is unclear. Possibly,
the tight binary simply forms via thermal fragmentation,
just like C1-Sa1 and C1-Sa2 in C1-Sa. Then, 
it remains to be seen how the disk angular momenta are
determined at $\la$100 AU scale. 

On the other hand, we can assume that  the source which used to drive
C1-Sa outflow2 was born first, and the other source that drives
C1-Sa outflow1 formed later. We assume this  order simply because
C1-Sa outflow2 is much longer (in size) than C1-Sa outflow1
(unless C1-Sa outflow1 is oriented close to the line-of-sight).
Based on the rough estimation in \S\ref{subsec:ld}, the accretion
associated with C1-Sa outflow2 was drastically reduced about 770 yr ago.
But about 3 kyr (\S\ref{subsec:out2}) ago, the accretion associated 
with C1-Sa outflow1 began (\S\ref{subsec:outflows}).
This time mismatch argues against the
scenario in which a protostellar outflow changes orientation
due to turbulent accretion \citep[e.g.,][]{2016ApJ...827L..11O}.
So the formation history of the tight binary
could be described as the following: a protostar formed in C1-Sa1 
first and launched C1-Sa outflow2; then another protostar formed
and launched C1-Sa outflow1; both protostars were 
accreting for some time before the first protostar, which
was driving C1-Sa outflow2, switched to a much weaker accretion. 

The fragmentation of C1-Sa2 provides a valuable example of
the primordial state of binary formation (if further confirmed
with a better sensitivity). Although currently
we have no kinematic information for C1-Sa2, the fact that
it appears as a spherically symmetric structure
(Figure \ref{fig:showbinary}(d), Figure \ref{fig:showout}(c)(d)) 
similar to C1-Sa1
indicates that it is a bound entity. So the pre-binary in
C1-Sa2 is also likely bound. The boundness of the final binary 
system depends on the accretion. The fragmentation
length ($\sim130$ AU) and mass ($\sim$0.17 M$_\odot$)
are larger than the Jeans scale ($\sim57$ AU)
and Jeans mass ($\sim$0.004 M$_\odot$) of C1-Sa2.
So perhaps further fragmentation will happen.
Note, at such high density of C1-Sa2, the free-fall timescale is 
just $\sim$550 yr. Once the protostar lights up,
the temperature increase should suppress fragmentation
\citep[e.g.,][]{2006ApJ...641L..45K}. 

\subsection{Comparison with other observations and theories}\label{subsec:compare}

Compared with other protobinary observations, C1-Sa is in a very
different environment. It is in the highly extincted, quiescent region
of the massive Dragon IRDC
\citep[the southern Dragon Tail,][]{2016ApJ...829L..19L,2019ApJ...873...31K}. 
There is no nearby massive star/protostar
within 5 pc. This condition is quite different from, e.g., 
IRAS 16547-4247 \citep{2020ApJ...900L...2T}, 
or G11.92-0.61 MM2 \citep{2022ApJ...931L..31C},
both having in-situ/nearby massive protostar.
Therefore, C1-Sa is one of the first-generation stars in the IRDC.
Its environment has not been impacted by fierce feedback from massive stars.
T16 estimated the mass outflow rate and the momentum injection rate for
C1-Sa outflow1, and concluded that C1-Sa is a good candidate for early-stage
massive protostar. Given now we show that C1-Sa is a binary system,
it provides an excellent example of the early, pristine stage of
massive binary formation.

If the C1-Sa fragmentation (which gives C1-Sa1 and C1-Sa2) is via disk 
fragmentation, we should see disk/spiral structures in the continuum map 
\citep[e.g.,][]{2009Sci...323..754K,2020A&A...644A..41O,2021A&A...652A..69M}. 
At 20 K, the column density detection limit 
(again using dust continuum) is $\sim2.0$ g cm$^{-2}$ or 
$N_{\rm H_2}=4.4\times10^{23}$ cm$^{-2}$, which should allow us
to see spiral structures similar to those in the above simulations.
If we assume a higher temperature as in C1-Sa1, the detection limit
becomes even lower. One could argue that C1-Sa is edge-on so that the 
elongation in C1-Sa is the disk plane. However, the C$^{18}$O 
position-velocity diagrams (\S\ref{subsec:ld}) show no obvious 
velocity gradients across C1-Sa. Possibly, the structures are 
rather smooth and not captured by the interferometric images.
The C1-Sa1 and C1-Sa2 separation matches well with that in
\citet{2009Sci...323..754K} though.

If the C1-Sa fragmentation is simply a thermal fragmentation,
it is still crucial to know what exactly the elongated structure 
is in C1-Sa. It shows little velocity gradient, 
as illustrated in the C$^{18}$O 
position-velocity diagrams (\S\ref{subsec:ld}). It aligns well with
magnetic fields \citep{2020ApJ...895..142L}. The field appears to be quite 
uniform (to be confirmed with more observations),
suggestive of a sub-Alfv\'enic condition, which disfavors
binary formation \citep[e.g.,][]{2020AJ....160...78R,2021A&A...652A..69M}.
If so, the field needs to dissipate with certain non-ideal MHD effects.
The elongated structure shows multiple small continuum peaks that are 
either interferometric effects or local fragments. 

The hypothesized binary in C1-Sa1 has a separation below 96 AU.
Their outflows appear to be quite straight and collimated, showing
no clear sign of precession, suggesting that the relevant accretion
has been quite stable. The binary members probably have formed in-situ
and their separation reflects their fragmentation scale. Such a small
scale favors the disk fragmentation scenario rather than turbulent
fragmentation \citep{2022arXiv220310066O}. However, their misaligned
outflows argue against disk fragmentation. Higher angular resolution
observations are necessary to show the sub-structures of C1-Sa1.

A bolder hypothesis is that C1-Sa1 evolved just like C1-Sa2,
but earlier, since the two kernels are born in the same environment.
So the structure of C1-Sa2 gives us clues of what happened in
C1-Sa1. The two fragments in C1-Sa2 roughly align on a 
filamentary structure, although C1-Sa2 itself is approximately
spherical at a larger scale. Once one of the two fragments in C1-Sa2
forms a protostar, which it may already be doing, the entire C1-Sa2
kernel is lit up. The two fragments may possess
gas with different angular momentum and thus launch 
misaligned outflows. 

\section{Conclusion}\label{sec:conc}

In this paper, we have reported ALMA Cycle 7 high-resolution observations toward the massive protostellar core C1-Sa, located at the highest extinction region in the Tail of the Dragon IRDC (aka G28.34+0.06 or G28.37+0.07). At the resolution scale of 140 AU, C1-Sa fragments into two centrally-peaked kernels (C1-Sa1 and C1-Sa2), each having a diameter of $\sim$420 AU. Their projected separation is $\sim$1400 AU, comparable to the Jeans scale of 1100 AU, consistent with a thermal fragmentation origin, unless the young binary is highly inclined to the line-of-sight. The thermal fragmentation is also consistent with the fact that C1-Sa has a sub/transonic non-thermal velocity dispersion based on the C$^{18}$O modeling. We suspect the non-thermal motion (partially) comes from gravitational collapse because the C$^{18}$O linewidth increases toward smaller scales. 

There are two bipolar outflows coming out of C1-Sa, which we name  C1-Sa outflow1 and C1-Sa outflow2. C1-Sa outflow1 was previously identified in T16 and K18 while C1-Sa outflow2 is only confirmed in this work with CO and SiO probably because C1-Sa outflow2 is parallel to the plane-of-the-sky. A detailed inspection suggests that both outflows are launched by C1-Sa1, the southern protostellar kernel in C1-Sa. C1-Sa2 is likely starless, which is consistent with the presence of N$_2$D$^+$ and ortho-H$_2$D$^+$ in its vicinity. There is a gap between C1-Sa outflow2 and C1-Sa1, indicating that the outflow was significantly reduced about 770 yr ago. C1-Sa outflow1 shows at least 6 bursty shocks, with interval between shocks of order 500 yr. We argue that C1-Sa1 has a tight binary, with each member launching a bipolar outflow and so presumably having its own accretion disk. Based on the large angle between C1-Sa outflow1 and C1-Sa outflow2, we argue that the two disks are misaligned.

C1-Sa1 is clearly detected in C$^{18}$O(2-1) while C1-Sa2 has no obvious line detection. The latter is consistent with C1-Sa2 being starless. Based on RADEX modeling, the C$^{18}$O line in C1-Sa1 is optically thick. So the double-peak feature in the beam spectrum and the lack of red-shifted emission in the position-velocity diagrams are likely due to self-absorption from the nearside cold infalling gas at the center of the kernel. The excitation temperature for the C$^{18}$O line is about 75 K, which is also likely the gas and dust temperature in C1-Sa1 at the density of $n_{\rm H_2}=1.8\times10^9$ cm$^{-3}$. The mass of C1-Sa1 is thus 0.55 M$_\odot$, a factor of 3 smaller than the presumably starless C1-Sa2 kernel, which we attribute to the protostellar accretion in C1-Sa1.

At the extreme resolution of the ALMA data (96 AU), C1-Sa2 further fragments into two sub-kernels with equal masses of 0.17 M$_\odot$, which requires further confirmation with higher sensitivities. The two fragments have a separation of $\sim$130 AU which is more than twice the C1-Sa2 Jeans scale at 20 K. At a density of $\sim6.1\times10^9$ cm$^{-3}$, C1-Sa2 has a free-fall timescale of just $\sim550$ yr. A protostar could form before further fragmentation takes place. C1-Sa2 provides a good example of the initial condition for binary star formation. Meanwhile, C1-Sa1 shows deviation from the circular shape, hinting that it could further fragment,  consistent with our speculation of the existence of a tight binary in this kernel. However, the hypothesized binary is not resolved at 96 AU.

The binary system is one of the earliest-stage 
forming binaries observed so far, which can potentially become massive.
Its birth environment is a 100 $\mu$m-dark, potentially strongly 
magnetized massive core that is
well-shielded and relatively quiescent (not impacted by nearby massive
stars). The high surface density of this IRDC  
also creates a high pressure environment for the deeply embedded binary 
system. 
Therefore, the C1-Sa binary system gives us the opportunity to
study (massive) binary formation in a very special environment.
More similar studies will show how binary formation
in such environments in IRDCs differs from binaries in other clouds.
To our knowledge, the C1-Sa binary
system is also the most distant forming binary observed thus far,
reaching $\sim$5 kpc toward the near end of the Galactic Bar.
Our study of the C1-Sa binary demonstrates that with ALMA, 
forming binaries can be detected and studied at these distances, 
significantly enlarging the parameter/volume space in which binary formation 
can be identified and studied, including a great number of IRDCs.

\acknowledgments 
We thank the anonymous referee for the constructive reports.
This paper makes use of the following ALMA data:
ADS/JAO.ALMA\#2013.1.00248.S,
ADS/JAO.ALMA\#2015.1.00183.S,
ADS/JAO.ALMA\#2016.1.00988.S,
and ADS/JAO.ALMA\#2019.1.00255.S.
ALMA is a partnership of ESO
(representing its member states), NSF (USA) and NINS (Japan), together
with NRC (Canada), NSC and ASIAA (Taiwan), and KASI (Republic of
Korea), in cooperation with the Republic of Chile.  The Joint ALMA
Observatory is operated by ESO, AUI/NRAO and NAOJ.  The National Radio
Astronomy Observatory is a facility of the National Science Foundation
operated under cooperative agreement by Associated Universities, Inc.
TGSP gratefully acknowledges support by the National Science Foundation under grant No. AST-2009842 and AST-2108989. 

\software{Python \citep{python}, SciPy \citep{scipy}, Astropy \citep{Astropy-Collaboration13}, Numpy \citep{numpy}, Matplotlib \citep{matplotlib}, SAOImageDS9 \citep{2003ASPC..295..489J}, CASA \citep{2022PASP..134k4501C}, RADEX \citep{2007A&A...468..627V}}

\facility{ALMA}

\appendix
\restartappendixnumbering 

\section{Distance to the Dragon IRDC}\label{sec:dist}

The kinematic distance to the Dragon Nebula is $\sim$4.8 kpc \citep{1998ApJ...508..721C}. The cloud resides in the first Galactic quadrant at l$\sim$28 degrees where the radial velocity will yield a near and a far distance solution for kinematic distance measurements. Since the cloud is seen in absorption in the infrared against the diffuse emission of the galactic plane, a near distance has been widely adopted. HI self absorption (HISA) analysis towards this cloud shows a pronounced absorption feature at the systemic velocity \citep{2010ApJ...721.1319A}. This adds further confidence to the choice of the near distance. Depending on the model for the galactic rotation curve \citep[e.g.,][]{1985ApJ...295..422C,2009ApJ...700..137R}, a distance between 4.5 and 5.0 kpc has been reported in the literature \citep{1998ApJ...508..721C,2006ApJ...653.1325S,2012A&A...547A..49R}.

Since the previous IRDC distance estimates, the galactic rotation curve \citep{2014ApJ...783..130R} along with the revised solar motion parameters \citep{2019ApJ...885..131R} have been updated. We have therefore used the \href{http://bessel.vlbi-astrometry.org/revised_kd_2014?}{online BeSSeL parallax based Distance Calculator} that combines kinematic distance information with displacement from the galactic plane and proximity to individual parallax sources from the BeSSeL parallax survey \citep{2019ApJ...885..131R} to generate a more complete distance probability density function for the cloud's distance assignment. The systemic velocity of the cloud based on published H$_2$CO and NH$_3$ measurements is $\sim$78 km s$^{-1}$ \citep{1998ApJ...508..721C,2006A&A...450..569P}. The  estimate of the cloud distance we obtain using this procedure is $4.34\pm0.30$ kpc. Taking all the above information into consideration, we adopt 4.8 kpc as the fiducial distance to the IRDC to be consistent with previous literature.

\section{ortho-H$_2$D$^+$ channel map}\label{sec:oh2dpchanmap}

\begin{figure*}[htb!]
\centering
\epsscale{1.1}
\plotone{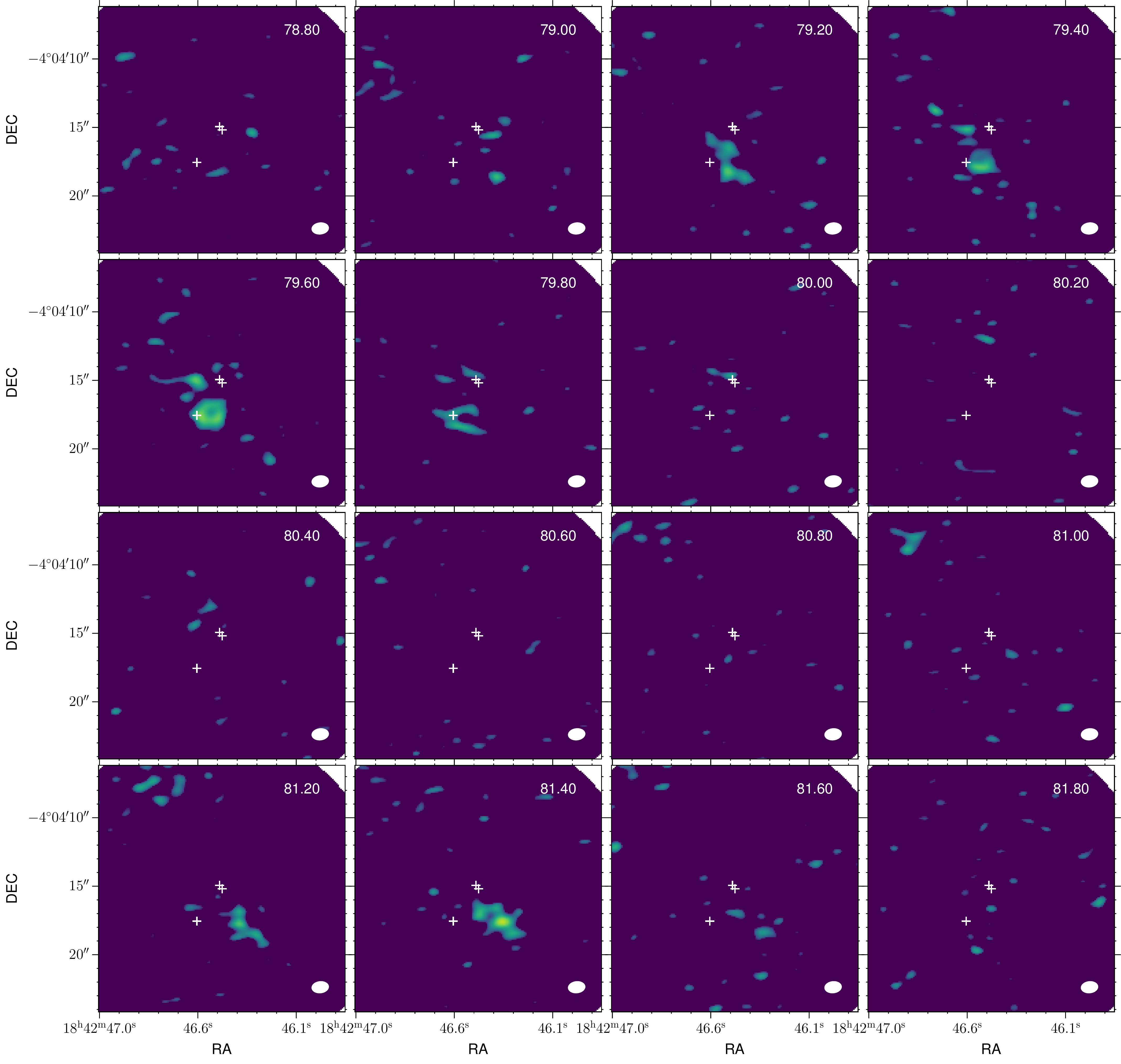}
\caption{
ortho-H$_2$D$^+$ channel maps with detection. The channel velocity is labeled at the top-right in km s$^{-1}$. The two kernels are marked as white crosses (C1-Sa1 at the map center). C1-Sb core is also marked (to the south-east from C1-Sa1). Only emission at SNR$\geq$2 is shown.
\label{fig:chanoh2dp}}
\end{figure*}

Figure \ref{fig:chanoh2dp} shows the ortho-H$_2$D$^+$ channel maps from the {\tt C4} data. The velocity range encompasses the detected ortho-H$_2$D$^+$ emission, which overlaps with the N$_2$D$^+$(3-2) emission detected in K18. From 79.2 km s$^{-1}$ to 79.8 km s$^{-1}$, the ortho-H$_2$D$^+$ emission lies to the southeast of the two kernels, with the peak emission in the 79.6 km s$^{-1}$ channel. The location of the peak emission coincides with the N$_2$D$^+$ core C1-S in K18. The high level of deuteration indicates that C1-S is at an early stage. One possibility is that C1-Sa, C1-Sb, and C1-S once belonged to a larger N$_2$D$^+$ core \citep{2013ApJ...779...96T}. Once C1-Sa and C1-Sb became protostellar, their level of deuteration decreased, leaving over the C1-S core which can develop into another protostellar system (K18). 

At 81.4 km s$^{-1}$, there is another peak emission of ortho-H$_2$D$^+$ to the south-west of C1-Sa, which was never noticed before. Perhaps it will develop into another core/protostar. Currently, it only has detection within 0.4 km s$^{-1}$ (two channels). The channel velocity coincides with the C1-N core from \citet{2013ApJ...779...96T}, which has a peak N$_2$D$^+$ emission at $\sim$81.2 km s$^{-1}$. The tentative detection at the top-left corner in channel 81.2 km s$^{-1}$ in Figure \ref{fig:chanoh2dp} spatially coincides with C1-N, which is at the highest surface density region in Dragon Tail.

\bibliography{ref}
\bibliographystyle{aasjournal}

\end{document}